# Scoping Review of AI, Metrology, and ESG in the Semiconductor Sector: Implications for Safe and Sustainable by Design (SSbD)

Karen Ang
Dept. Process Engineering
Infineon Technologies Sdn. Bhd.
Kulim, Malaysia
hueiling.karenang@infineon.com
0009-0008-5923-0106

Han-Teng Liao* (Corresponding Author)
*Independent Researcher*
Penang, Malaysia
h.liao@ieee.org
0000-0003-1081-5599

*Abstract*— **The semiconductor sector faces a dual transition: scaling manufacturing execution through Artificial Intelligence (AI) while satisfying stringent sustainability mandates, such as the EU Carbon Border Adjustment Mechanism (CBAM). This paper presents a scoping review of 1,465 documents indexed in Web of Science and Scopus, spanning AI-integrated metrology, supply chain ESG, and federated industrial data spaces. Network analysis reveals a highly fragmented "core-periphery" knowledge structure, emphasizing a critical structural hole between AI-driven process optimization and downstream sustainability governance. To close these gaps, this study proposes a 6-layer Safe and Sustainable by Design (SSbD) architecture grounded in a System of Systems (SoS) paradigm. By establishing distinct "grid-to-core" and "standards-through-supply-chain" integration pathways, the proposed framework demonstrates how virtual metrology (VM), localized federated learning, and defensive RegTech mechanisms can build provenance-aware data fabrics. Ultimately, this architecture positions regulatory compliance as a driver for innovation, enabling secure, climate-neutral, and circular value chains in semiconductor manufacturing.**

## I. Introduction

The semiconductor industry currently navigates a profound and high-stakes Twin Transition. On one hand, the Artificial Intelligence (AI) explosive demand requires rapid scaling and precision in advanced manufacturing. On the other, Environmental, Social, and Governance (ESG) mandates, epitomized by the European Union's Carbon Border Adjustment Mechanism (CBAM)—a climate measure on imported goods—and the Corporate Supply Chain Due Diligence Directive (CSDDD)—a supply chain accountability regulation—, now pressure the sector to scale processing throughput while minimizing footprints.

Such transition unfolds amid volatility. Public health and geopolitical crises have exposed structural vulnerabilities in the global network [1]. Systemic disruptions—such as the COVID-19-induced chip shortage—propagate rapidly across industries, creating prolonged shortages in sectors like automotive manufacturing [2]. Historical precedents, including the 2011 Tohoku earthquake, underscored the need to safeguard the supply of microcontroller units [3]. Recent geopolitical struggles, linking AI chips and rare earth minerals, impact other sectors such as automotive, energy, defence, aerospace, and AI data centres [4].

Ensuring semiconductor supply chain resilience has become a global priority, necessitating integrated frameworks that coordinate strategic pricing, demand, and capital expenditure. As fabrication moves towards sub-5 nm nodes, the 2024 International Roadmap for Devices and Systems (IRDS) Environment, Safety, Health & Sustainability: Environmental Sustainability of the Semiconductor Facilities (ESHS-ESSF) defines a technical roadmap for sustainable manufacturing [5]. Concurrently, advances in Virtual Metrology (VM) [6], deep learning [7], and predictive models [8] enable real-time monitoring for Statistical Process Control (SPC) —a method for monitoring production— in modern fabrication [9]. Initiatives like the **imec.netzero** "virtual fab" platform, launched by the Interuniversity Microelectronics Centre (IMEC), extends such predictive capability by quantifying energy and environmental footprints.

### A. Motivation

How might we leverage the energy and emission data insights into design, facilitating broader regulatory compliance for various stakeholders? The EU-initiated Safe and Sustainable by Design (SSbD) chemical safety framework [10], [11] is expected to drive this industrial transformation, ensuring material safety for both human health and the environment. To adapt, companies must navigate the evolving regulatory landscapes of the CBAM, SSbD and CSDDD with a Governance, Risk, and Compliance (GRC) strategy [12]. This approach allows them to manage material and product lifecycles in alignment with UN Sustainable Development Goals 9 (Industry and Innovation) and 12 (Responsible Production). Crucially, it synchronizes high-precision manufacturing with transparent emissions reporting, advancing corporate ESG practices from basic administrative accountability to science-based, empirical evidence.

This integration is especially critical for the semiconductor sector, which is inherently intensive in its use of rare earth minerals, scarce materials, and hazardous chemicals. Embedding sustainability norms directly into procurement and production workflows via digital technologies enables the sector to pursue system-wide lifecycle orchestration that is both technologically advanced and ecologically responsible [13]. Emerging Regulatory Technology (RegTech)—digital tools for compliance monitoring—service ecosystems operationalize this integration by evaluating suppliers, automating compliance screening, and tracking process emissions. Positioned at the material-financial interface of manufacturing, these systems constitute a foundation for ESG data integration, within the "chemical-to-chip" value chain.

TABLE I. SELECTIVE 2026 REGTECH PLATFORMS AND STANDARDS ALIGNMENT FOR SEMICONDUCTOR MANUFACTURING

| Platform | Core Functionality | Standards Alignment | CBAM/ESG Relevance | Critical Assessment |
|---|---|---|---|---|
| **EcoVadis** | Supplier ESG ratings, Scope 3 decarbonization | Global Reporting Initiative, ISO 26000, SBTi | Strong on procurement ESG, limited ITU/IEEE linkage | Widely adopted, lacks semiconductor specificity |
| **IntegrityNext** | Supplier sustainability and compliance automation | Integrates with SAP Ariba, ISO risk indicators | Supports ESG risk detection, CBAM readiness | Strong orchestration, but limited transparency on standards |
| **Workiva** | ESG reporting, audit-ready compliance | CSRD, ESRS, ISSB, GRI, and California climate laws | Strong CBAM disclosure potential | Focused on reporting, less on semiconductor process data |
| **AuditBoard** | Risk & compliance management | SOC 2, ISO, NIST, PCI, COSO, HIPAA, etc. | Supports governance, CBAM audit | Governance-centric, limited metrology integration |
| **SAP Ariba** | Procurement & supplier management | ISO | Strong ESG procurement, CBAM readiness | Enterprise-scale, but limited SSbD lifecycle focus |

Note. Based on rapid scoping of publicly available information. As ESG service ecosystems are dynamic, the assessments may become outdated and require periodic re-validation.

Still, their alignment with key international standards (ITU-T L.1470/L.1480, IEEE 7000, ISO 14000 series, and Science Based Targets initiative) and regulatory frameworks remains uneven and platform-dependent. This service ecosystem landscape reflects a deeper architectural deficit. Treating ESG platforms as discrete compliance tools obscures their potential role within interoperable, lifecycle-oriented systems. Semiconductor manufacturing, given its material complexity and global regulatory exposure, demands architectures informed by SSbD principles —ones that transcend departmental and platform silos by integrating AI-driven analytics, precision metrology, and sustainability mandates into coherent governance frameworks. This gap motivates the present scoping review, which surveys standards, metrology, and federated data spaces as foundational enablers for green digital ecosystem innovation.

### B. *Objectives*

The central guiding question is: How can a scoping review inform the SSbD-aligned architectures integrating AI, metrology, and ESG compliance in semiconductor manufacturing? Three objectives structure this inquiry:

Objective 1 –**Integration:** Motivate a structured scoping of how semiconductor analytics, metrology, and ESG compliance can be connected into interoperable, lifecycle-oriented architectures.

Objective 2– **Constraints and Gaps:** Describe the challenges of scarce data, fragmented ESG reporting, and uneven platform alignment as a foundation for subsequent methodological questions on adaptation and reliability.

Objective 3 –**System-Level Efficiency and Assurance:** Argue that yield and throughput optimization, AI safety, and CBAM compliance are interconnected governance challenges requiring lifecycle-level treatment, not siloed technical fixes.

Collectively, these objectives justify the need to scope connections across metrology, AI, and ESG compliance

## II. EXISTING THEORIES & PREVIOUS WORK

Semiconductor manufacturing can be understood through multiple ecosystem lenses. The section describes related ecosystem gaps so as to integrate them into a System of Systems (SoS) challenge [14] to be synthesized into the SSbD architecture that will be proposed in Section V.

### A. *Supply Chain Material Safety and Phased Transitions*

The 2024 IRDS ESHS-ESSF roadmap sets the horizon, identifying PFAS, hazardous waste, and Scope 3 supply chain accountability [5]. The roadmap highlights phased transitions: near-term waste reduction, mid-term substitution under European Union Registration, Evaluation, Authorization and Restriction of Chemicals (EU REACH), and longer-term Scope 3 accountability, supported by facility-level water, energy, and ESG monitoring frameworks. Relevant for phased transitions, an integrated framework of Industry 3.5 was proposed for semiconductor manufacturing, suggesting a phased migration toward Industry 4.0 rather than abrupt one, showing incremental transformation viability under constraints [15]

While systematic literature exists for integrated supply chain production planning [16] and resilience [1], [17], [18], [19], no formal mechanism exists to propagate material safety outcomes into downstream compliance or design governance.

### B. *Ecosystems of Policy and RegTech Compliance*

Science-mapping of digital transformation in business ecosystems establishes Green Digital Transformation (GDT) roadmaps integrating natural, business, and digital ecosystem services as the governance frame for sustainable industrial transitions [20].. The broader shift from market-led to governance-led semiconductor ecosystems [21], driven by geopolitical interdependence and supply chain vulnerability, has been well documented [22], [23], [24]. CBAM (operational 2026) creates indirect carbon cost exposure for semiconductor supply chains [25].

While digital transformation is increasingly linked to improved ESG reporting performance under mandatory disclosure regimes [26], no RegTech compliance architecture yet bridges process-level material data with financial and regulatory accountability in the sector, including the ESG reporting orchestration services (e.g., IntegrityNext).

### C. *Federated Data Governance and Interoperability*

Cross-data space interworking has proven feasible for $CO_2$ footprint monitoring, between Europe's International Data Spaces (IDS) and Japan's Cross-domain Architecture for Data Exchange (CADDE) frameworks, showing that sovereign data exchange can support accountability across regions, overcoming technological and legislative boundaries [27]. For semiconductors, federated data spaces could integrate metrology, ESG platforms, and RegTech tools, using interoperable architectures. Embedding VM and defect detection pipelines into federated data fabrics would align technical monitoring with CBAM and SSbD compliance, enabling system-wide accountability.

### D. *Metrology, Smart Manufacturing, and Optimisation*

AI-enabled metrology boosts yield and reliability. VM and defect detection methods [6], [13] use predictive analytics to minimize variability and boost throughput. Smart sensors and closed-loop optimization extend these gains, embedding measurement into real-time process cyber-physical systems [28]. Yet these advances remain siloed without lifecycle governance. Merging metrology with ESG and CBAM data flows advances sustainability beyond production efficiency.

### *E. Industry 5.0 and Socio-Technical Transition*

Industry 5.0 reframes manufacturing around human-centric, resilient, and sustainable value creation beyond efficiency alone [29], including using foundation and multimodal models [30], [31]. Still, it remains an open question how the industry applications of the AI models contribute to the socio-technical transition [32], [33].

### *F. A System of Systems (SoS) challenge*

Semiconductor manufacturing needs systems thinking of lifecycle assessment (LCA) and circular economy principles, addressing hazardous waste, and energy consumption [5]. From the **socio-technical systems** perspective—an Industry 5.0 approach integrating social governance with industrial technology—, digital product passports and circular economy architectures operationalize Industry 5.0 by tracing product lifecycles and supporting human–machine synergy and sustainability [34]. While standards such as ITU-T L.1470/L.1480, IEEE 7000, and ISO 14000 provide governance scaffolding [35], but few studies integrate them.

By framing metrology, AI analytics, and ESG compliance as interdependent services, the study contextualizes SSbD as a **System of Systems (SoS)** —a complex integration of interoperable systems— challenge, underscoring the need for federated data, interoperable governance, and architectures to achieve safe, climate-neutral, and circular innovation.

## III. METHODOLOGY: SCOPING FOR CONNECTIONS

To better integrate the fragmented but complementary strands of literature, this section describes a scoping review designed to bridge structural knowledge gaps and conceptual framework, and also integrate research fronts into a lifecycle-oriented SSbD architecture.

### *A. Study Scope and Research Questions*

This study translates engineering challenges into technology management and systems engineering perspectives through three research questions:

RQ1 -- **Integration**: How can semiconductor analytics and measurement systems integrate heterogeneous data and generalize across customers and process nodes while remaining interoperable with ESG frameworks, including through federated data space architecture integration?

RQ2 -- **Adaptation Under Constraints**: How can analytics and measurement platforms be designed and tuned for scarce or imbalanced data while maintaining reliability for critical defect classes, leveraging soft-sensing, VM, and outsourced pipelines for better resource allocation, risk management, and lifecycle assurance?

RQ3 -- **Efficiency and Compliance**: How can industrial analytics improve yield and throughput while preserving performance for underrepresented classes and aligning with AI safety and sustainability mandates (e.g., Scope 3, CBAM) -- framing optimization as a SoS challenge rather than a purely algorithmic one? This SoS approach underscores the necessity for integrated data architectures to achieve safe, climate-neutral, and circular innovation..

### *B. Applied Research Methods*

To answer these questions, we combined a scoping review with technology roadmapping on selected six thematic domains (see Table II), which roughly follows the six areas covered in Section II. Enterprise GRC platforms—such as EcoVadis, IntegrityNext, and SAP Ariba are also included.

TABLE II. SIX DOMAINS AND THEIR SEARCH QUERIES AND RECORD COUNTS (WOS/SCOPUS)

| Domain | Rationale | Query Logic (WoS syntax shown, TS=) [a] | N (WoS) | N (Scopus) |
|---|---|---|---|---|
| 1. Supply Chain and Emissions | Analyzes semiconductor supply chains with Scope 3 emissions, carbon neutrality, resilience, and safety, aligning sustainability and industrial goals. | ( (semiconductor* OR "Wafer Fab*" ) AND ( "supply chain" ) AND ("Scope 3" OR "carbon neutrality" OR "carbon emissions" OR resilience OR safety) ) | 121 | 237 |
| 2. Ecosystems of Policy and RegTech (ESG Platforms) | Explores semiconductor and digital platforms aligned with ITU, IEEE, ISO, CBAM, and ESG rating systems, advancing responsible business alliances and green digital transformation. | ( (semiconductor* OR "Wafer Fab*" OR "digital platform*" ) AND ("ITU-T L.1470" OR "ITU-T L.1480" OR "IEEE 7000" OR "IEEE standard*" OR "ITU standard*" OR "ISO standard*" OR "green standard*" OR "Carbon Border Adjustment Mechanism" OR "Responsible Business Alliance" ) OR (EcoVadis OR IntegrityNext OR Workiva OR AuditBoard OR "SAP Ariba") )[b] | 52 | 260 |
| 3. Federated Data Governance and Interoperability | Addresses federated data spaces and interoperability, enabling secure, standards-aligned digital collaboration for trustworthy platforms. | ( (semiconductor* OR "Wafer Fab*" ) AND ("federated data" OR "data space*" OR interoperability) ) | 45 | 285 |
| 4. Metrology and AI Integration | Integrates metrology and in-situ sensors with AI, machine learning, and analytics, enabling measurement, and process optimization. | ( (semiconductor* OR "Wafer Fab*" ) AND ( metrology OR "In-situ sensor*" ) AND ("artificial intelligence" OR "machine learning" OR "data analytics" )) | 99 | 265 |
| 5. Foundation and Multimodal Models | Examines foundation and multimodal models in semiconductor manufacturing and metrology, for safe, explainable, standards-driven intelligence. | ( (semiconductor* OR "Wafer Fab*" ) AND ( (multimodal OR foundation*) NEAR model* ) AND (manufacturing OR metrology ) ) | 35 | 14 |
| 6.Circular Economy and AI | Investigates semiconductor lifecycle through digital circularity, zero-waste engineering, and LCA, leveraging AI, machine learning, and analytics for sustainable manufacturing innovation. | ( (semiconductor* OR "Wafer Fab*" ) AND ( "Digital Circular" OR "Circular Economy" OR "Circular Engineering" OR "zero-waste" OR "life cycle" OR LCA ) AND ("artificial intelligence" OR "machine learning" OR "data analytics" )) | 17 | 47 |

a. Full Boolean queries (WoS and Scopus syntax) are provided for reproducibility. Scopus queries are identical to those shown, with TS= replaced by TITLE-ABS-KEY.

b. Specific SaaS providers (e.g., **IntegrityNext, SupplyShift, AuditBoard**) are included to identify literature at the intersection of "Software-as-a-Service" and "Sustainable Procurement." Their inclusion ensures the scoping review accounts for how theoretical ESG standards are practically implemented through automated risk management and supplier governance ecosystems. These tools represent the "operational layer" where semiconductor firms manage vendor compliance, Scope 3 data collection, and alignment with frameworks like CBAM.

### C. Technology Roadmapping Innovation Ecosystems

The scoping review findings inform a Technology Roadmapping (TRM) exercise [36], [37], to situate EU SSbD framework within semiconductor manufacturing processes. To operationalize this, the study applies the International Telecommunication Union (ITU) ICT-centric innovation ecosystem toolkits [38], which provide stakeholder mapping methods and co-creation canvases. These tools help translate bibliometric knowledge and stakeholder clusters into actionable pathways, ensuring SSbD principles are integrated into innovation ecosystems rather than treated as compliance burdens. The synergy between scoping review-supported TRM and ITU toolkits will then provide the basis for the study to propose a layered SSbD architecture for the industry, thereby strengthening capacity for academic research, industrial application, and regulatory compliance in the pursuit of climate-neutral and circular innovation.

### D. Data Collection

Targeted queries in Web of Science (WoS) and Scopus yielded 1,465 total documents in March 2026 (see Table III). To ensure reproducibility while maintaining compliance, the processed bibliometric data is available via the "osf_dataset-Semi_SSbD" repository (DOI: 10.5281/zenodo.20283923) selectively, including both the thesaurus and cleaned data to support reuse and replication [39].

TABLE III. DATASETS OVERVIEW

| Category | Main Information About Data | | |
|---|---|---|---|
| | ***Description*** | ***WoS*** | ***Scopus*** |
| Literature | Documents | 363 | 1102 |
| | Timespan | 1997~ 2026 | 1965~ 2026 |
| | Sources (Journals, Books, etc.) | 202 | 649 |
| | Total References | 20013 | 79506 |
| Growth & Freshness | Annual Growth Rate (%) | 4.22 | 6.28 |
| | Document Average Age | 4.86 | 8.19 |
| Impact | Average Citations Per Doc | 19 | 13.25 |
| Document Contents | Author's Keywords (ID) | 1636 | 3189 |
| | Keyword Plus (DE) | 829 | 8949 |
| Collabora-tion | Authors | 1558 | 4195 |
| | Co-Authors Per Doc | 4.61 | 4.5 |
| | International Co-authorships (%) | 26.72 | 20.87 |
| Document Types | Articles | 318 | 452 |
| | Conference Papers | 6 | 540 |
| | Books or Chapters | 2 | 35 |
| | Reviews or Surveys | 31 | 66 |
| | Misc. | 6 | 9 |

Both datasets confirm the growing and multidisciplinary nature of semiconductor research. WoS provides a highly curated, high-impact dataset (averaging 19 citations per document), optimized for mapping conceptual structures via author keywords or institutions; its author keyword data (DE tag) is significantly more complete, with less than 10% missing values compared to Scopus's 38%. Scopus captures a broader thematic spread and reflects active industrial knowledge cycles through 540 conference papers. While both reflect the codified knowledge transfer characteristic of the industry, high keyword specificity underscores domain fragmentation. This fragmentation motivates our scoping review to synthesize system-level frameworks that align technical innovation with governance requirements.

### E. Bibliometric Analysis and Visualization

To identify network structures and structural holes, bibliometric analysis was conducted using Bibliometrix (R-package)[40], VOSviewer, and customized Python scripts. Prior to analysis, synonyms were merged through an extensive data-cleaning to ensure taxonomic precision [41], [42] and clear conceptual mapping.

By examining both conceptual structure (keywords, thematic clusters) and the social structure (authors, institutions, countries) of the field, the scoping review surfaces critical structural holes across different research fronts —including metrology-AI integration, ESG compliance, and federated data spaces. These gaps inform the proposed SSbD architecture, establishing pathways to close these gaps by linking localized technical advances to system-wide socio-technical solutions for the semiconductor sector.

## IV. RESEARCH FINDINGS

### A. Affiliations: leading research and corporate institutions

Comparative analysis of affiliations from WoS and Scopus highlights distinct stakeholder patterns. The WoS dataset is dominated by universities and public institutions such as National Tsing Hua University, National Taipei University of Technology, Seoul National University, and the University of Texas System. National research organizations like the Chinese Academy of Sciences and CNRS also appear prominently, reflecting government-backed R&D shaping foundational knowledge. Industry presence is limited, with Samsung, Intel, and Seagate Technology as notable entries.

TABLE IV. TOP AFFILIATIONS

| Web of Science | | Scopus | |
|---|---|---|---|
| ***Affiliation*** | ***N*** | ***Affiliation*** | ***N*** |
| National Tsing Hua University | 13 | Infineon Technologies Ag | 26 |
| Chinese Academy of Sciences | 12 | Interuniversity Microelectronics Centre (IMEC) | 24 |
| University of Texas System | 10 | Chinese Academy of Sciences | 16 |
| National Taipei University of Technology | 9 | Samsung Electronics Co., Ltd. | 15 |
| State University System of Florida | 8 | Arizona State University | 15 |
| Arizona State University | 6 | National Tsing Hua University | 14 |
| Arizona State University Tempe | 6 | ASML Netherlands Bv | 14 |
| Myongji University | 6 | Intel Corporation | 13 |
| National Taiwan University | 6 | Centre National De La Recherche Scientifique (CNRC) | 12 |
| Samsung | 6 | Stmicroelectronics Sa, France | 12 |
| Seoul National University | 6 | Siemens Eda | 11 |
| University of Florida | 6 | Nova Measuring Instruments | 11 |

By contrast, Scopus emphasizes industry stakeholders. Affiliations such as Infineon Technologies, ASML Netherlands, Intel, Samsung Electronics, STMicroelectronics, Siemens EDA, and Nova Measuring Instruments dominate, underscoring the Scopus's broader institutional coverage of applied and corporate outputs. Collaborative hubs like IMEC also feature research bridging academia and industry. Universities remain visible—Arizona State University, National Tsing Hua University, etc.

Taken together, the datasets reveal complementary perspectives. WoS foregrounds academic and public ecosystems driving fundamental science and talent development. Scopus highlights industrial and collaborative R&D, capturing contributions of multinational corporations and cross-institutional centers translating research into practice. Both datasets include universities, national research institutions, and industry stakeholders, but differ in emphasis.

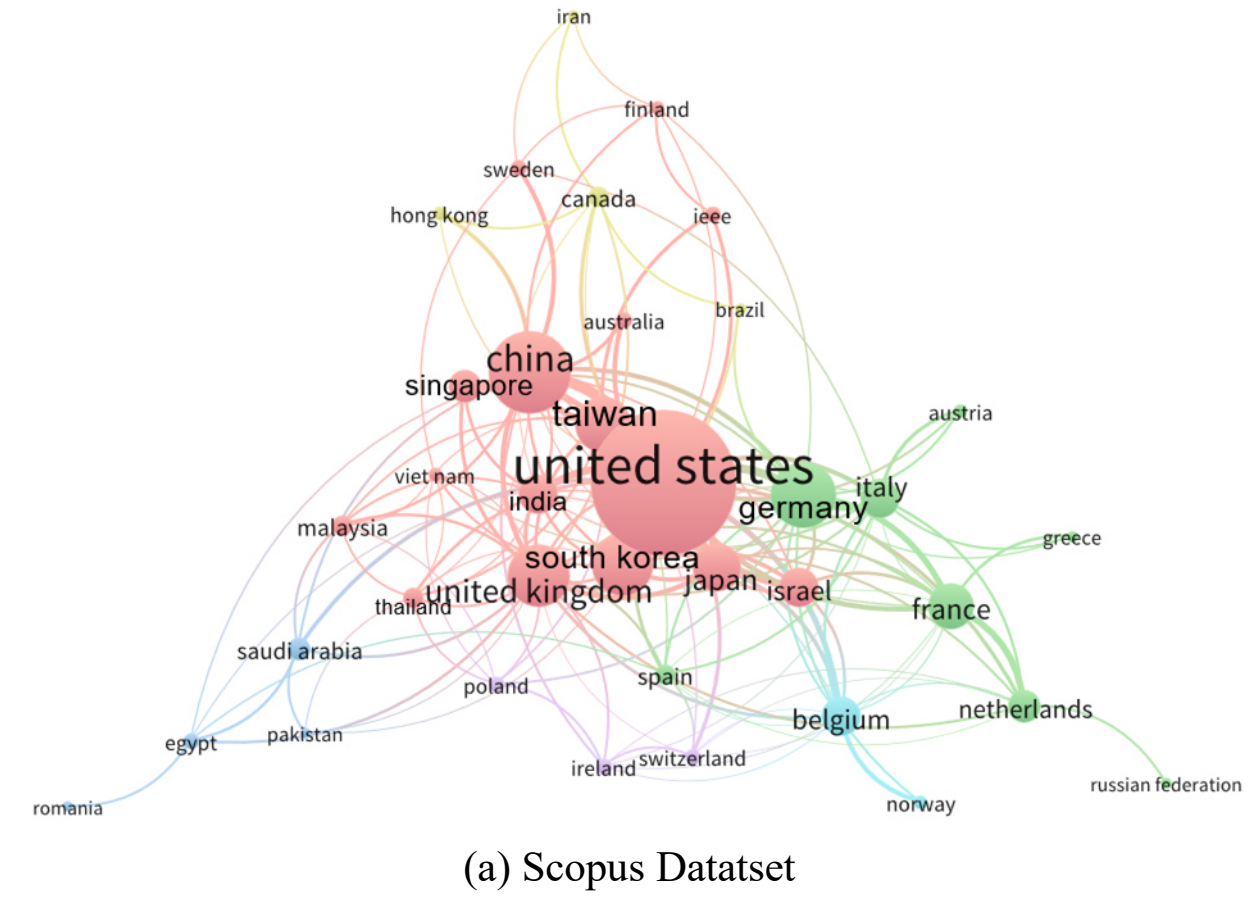

(a) Scopus Datatset

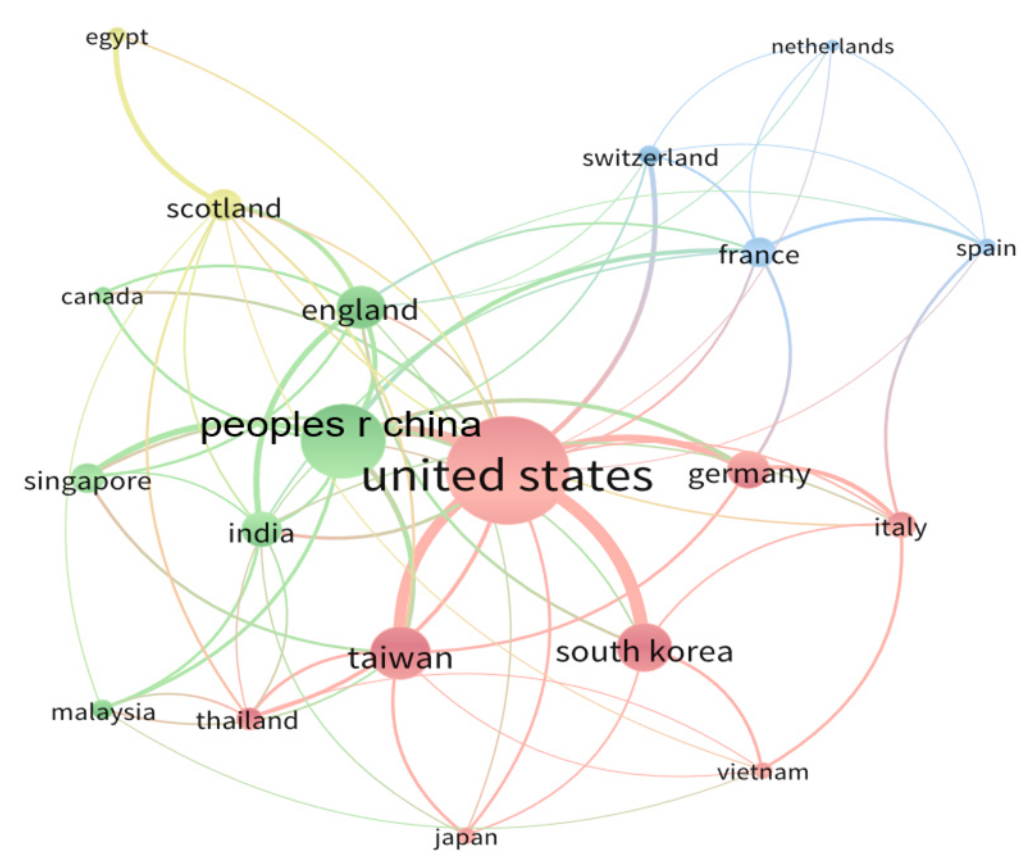

(b) Web of Science Datatset

Fig. 1. Top relevant countries: International Co-authorship

This divergence signals a structural hole — the absence of formalized knowledge exchange mechanisms linking public research outputs to industrial practices. Closing this gap calls for reference or consortium-based partnerships, such as IMEC-led multi-stakeholder platforms and IEEE IRDS roadmap taskforces, that institutionalize knowledge exchange among stakeholder ecosystems, such as fundamental research, standards organizations, professional associations, and corporate technical and sustainability compliance committees.

### *B. Countries: leading countries and collaboration patterns*

Co-authorship networks from Scopus and WoS (Fig. 1) reveal a highly integrated global landscape, where the United States, China, South Korea, and Taiwan serve as densely interconnected hubs with significant cross-border collaboration. Thick linkage lines and central positioning of these economies suggest that, despite geopolitical rhetoric, functional "decoupling" has not materialized in semiconductor research. Instead, the community remains characterized by strong trans-Pacific and Euro-Asian exchanges, underscoring persistent systemic interdependence in advancing manufacturing and supply chain resilience.

Yet this collaborative density masks a regulatory structural hole: shared research outputs are not matched by harmonized ESG reporting or data governance standards across jurisdictions. Bridging this gap requires multilateral alignment efforts—such as mutual recognition agreements between EU CBAM frameworks and East Asian national sustainability disclosure regimes— to ensure that research collaboration translates into interoperable compliance infrastructure.

### *C. Funding Agencies: leading funders and projects*

Table V provides an additional lens through funding agencies, within the WoS dataset.

TABLE V. TOP FUNDERS

| Region | Key Funding Agencies (Top Contributors) | N | % |
|---|---|---|---|
| China | NSFC, National Key R&D Program, CAS, Provincial Foundations | 81 | 21.77% |
| Non-China East Asia | NRF (Korea), MSIT, NSTC (Taiwan), JSPS (Japan), Samsung/TSMC | 78 | 20.97% |
| European Union | Horizon 2020/Europe, ECSEL JU, NextGenerationEU, ERC | 42 | 11.29% |
| United States | NSF, NIST, DoD, DoE, DARPA | 36 | 9.68% |

Note: Funding agency counts represent mentions. Because individual papers may acknowledge multiple agencies, totals exceed the 363 documents in the WoS dataset (N=372).

Leading contributors include the European Union (16 projects), the US National Science Foundation (16), National Research Foundation of Korea (14), Taiwan NSTC (11), and UK Research and Innovation UKRI (7), alongside corporate actors such as Samsung (5). When aggregated, East Asian agencies (Korea, Taiwan, China) account for 42.74% of the funded projects. This distribution underscores the distinct regional clustering of semiconductor R&D ecosystems and the critical interplay of public and corporate funding mechanisms in sustaining industry innovation.

The divergence between East Asian funding volume and Western citation impact points to a structural hole: high-output regional R&D ecosystems remain insufficiently connected to the governance frameworks that are shaping global market access requirements. Closing this gap would require dedicated co-funding instruments, such as EU–Asia bilateral research programs or IEEE-aligned technology roadmapping initiatives, that explicitly link funded research outputs to interoperable sustainability compliance pathways.

### *D. Conceptual structure*

The factorial map (Fig. 2), generated via Multiple Correspondence Analysis (MCA) on the WoS dataset, visualizes the relational structures across five clusters. Dim 1 (79.72%) represents the dominant axis of supply chains and digital transformation, while Dim 2 (88.71%) distinguishes between operational execution and strategic governance, marking a bifurcation between factory-floor optimization and SoS lifecycle orchestration.

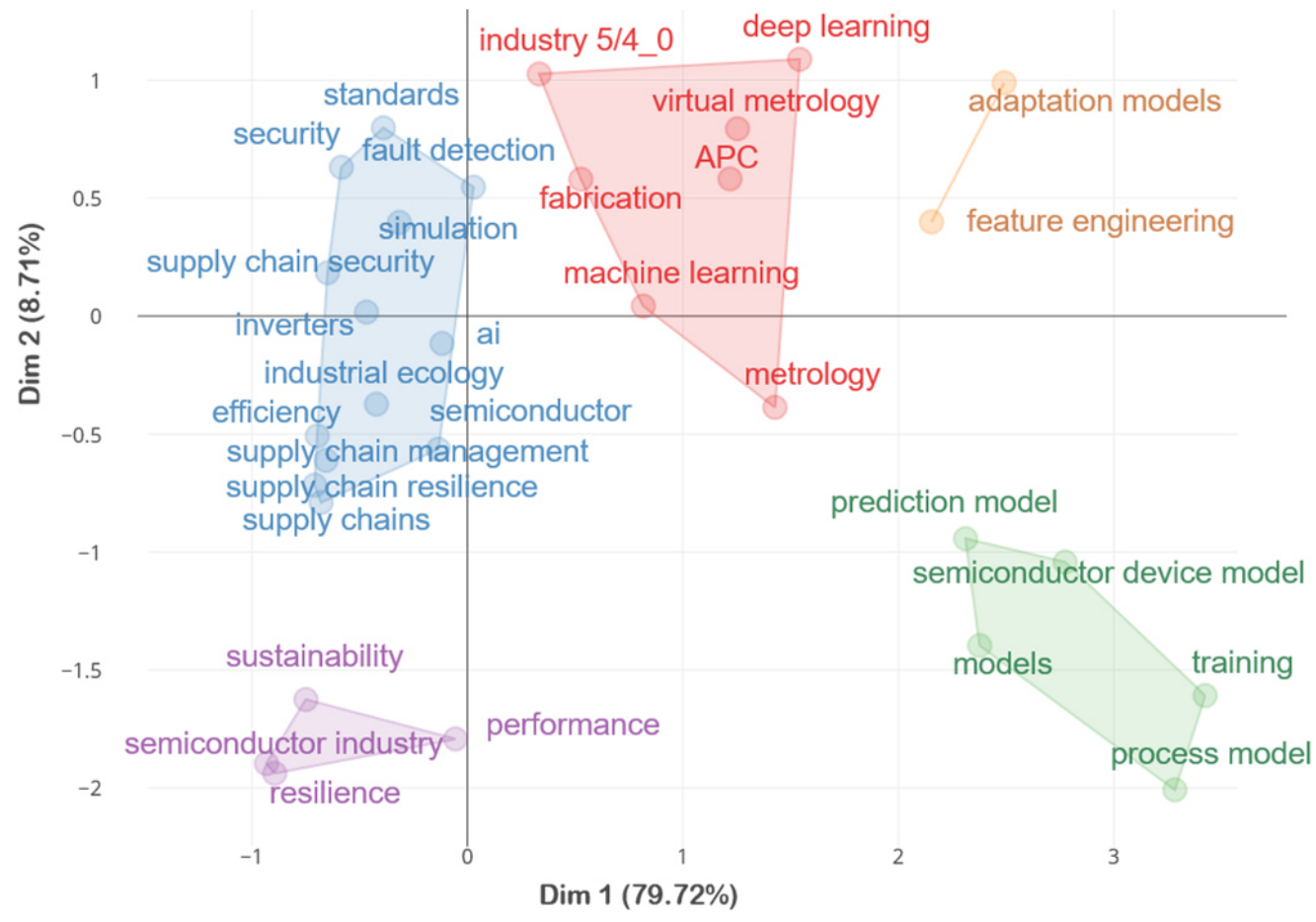

Fig. 2. Conceptual network: factorial (MCA) visualization

Cluster 1: *AI-Driven Smart Manufacturing* (Red): Anchored by deep learning, Virtual Metrology (VM), and Industry 4.0/5.0, this cluster defines the computational core of modern semiconductor fabrication. The tight co-occurrence of virtual metrology with deep learning signals a maturing research trajectory toward AI-native process control —a direct enabler of the SSbD metrology layer.

Cluster 2: *Supply Chain Resilience and Governance* (Blue): Grouping supply chains, standards, security, fault detection, inverters, and industrial ecology, this cluster reflects the strategic governance dimension of semiconductor manufacturing. The co-presence of security and standards alongside supply chains indicates a growing recognition that resilience is as much a regulatory and normative challenge as a logistical one.

Cluster 3: *Predictive and Process Modelling* (Green): Comprising process model, prediction model, models, adaptation models, and training, this cluster represents the model lifecycle layer — where predictive accuracy and model governance determine the reliability of risk-aware decision-making in production and compliance contexts.

Cluster *4: Sustainability and Device Performance* (Purple): Linking sustainability, performance, and semiconductor device modelling, this cluster is the smallest but strategically significant, positioning environmental accountability alongside physical device metrics. Its relative isolation from the AI-core cluster (Red) is a structural finding.

The most critical structural hole lies between Cluster 4 (sustainability and performance) and Cluster 1 (AI and virtual metrology): environmental impact metrics are not yet co-evolving with AI-driven process optimization. Bridging this gap requires embedding lifecycle assessment outputs and CBAM-relevant emissions data directly into virtual metrology feedback loops, so that sustainability performance is treated as an integrated control component alongside yield and throughput — a principle central to the proposed SSbD architecture.

To complement the factorial mapping, a keyword co-occurrence network (Fig. 3) visualizes thematic associations and the structural centrality of key terms. The network exhibits a pronounced core-periphery topology.

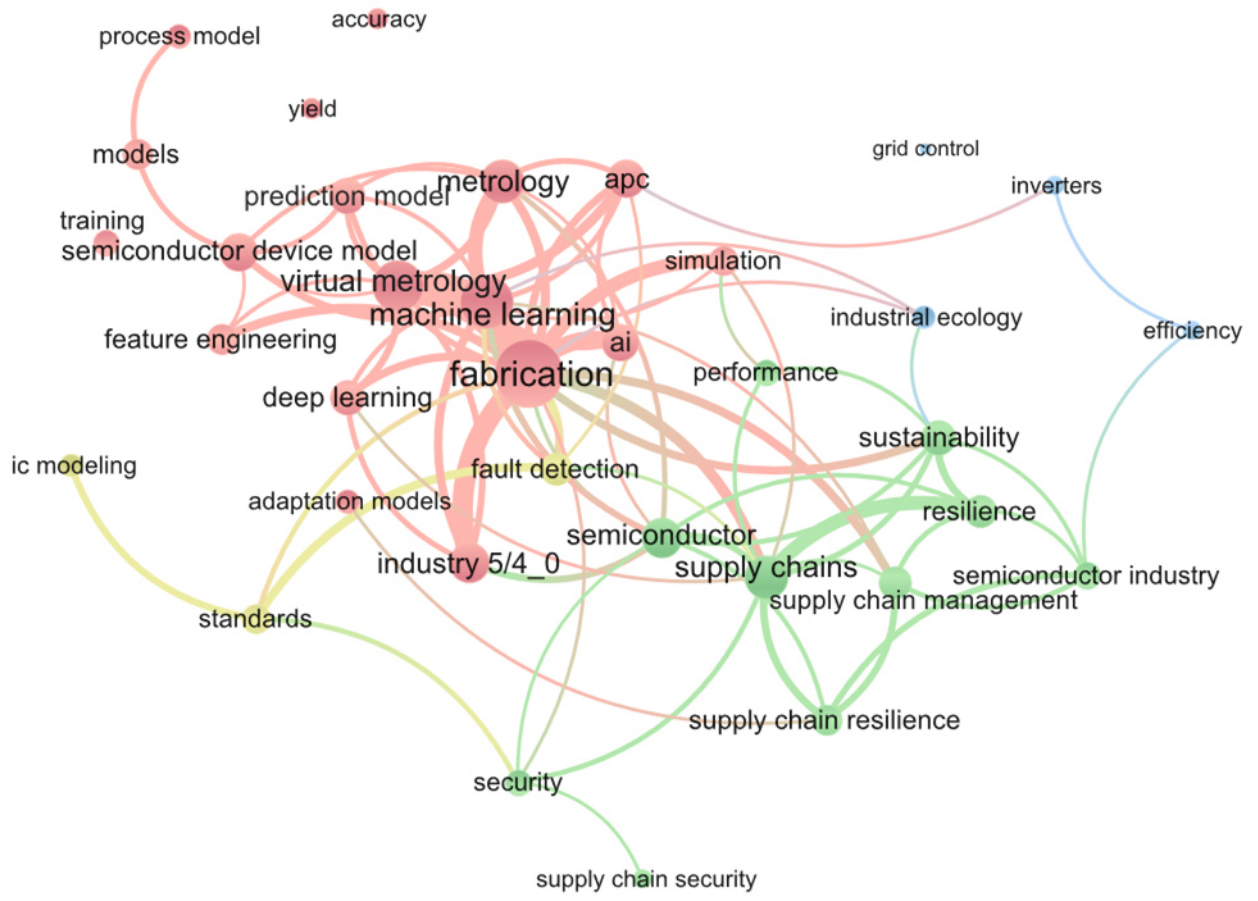

Fig. 3. Conceptual network: keyword co-occurrence visualization

**Cluster 1 — AI-Driven Fabrication Core (Red):** The densest and most central cluster, anchored by *fabrication, machine learning, virtual metrology*, and *metrology*, with tightly coupled satellites including *deep learning*, *prediction model*, *fault detection*, *APC*, *AI, simulation, yield, accuracy, training*, and *process model*. Its high density and central placement confirm AI-enabled process control as the dominant research paradigm.

**Cluster 2 — Supply Chain Resilience (Green):** A cohesive peripheral cluster grouped around *supply chain resilience*, *sustainability*, *semiconductor industry*, and *performance*. It signals a maturing sub-field.

**Cluster 3 — Standards and Governance (Yellow-Green):** A smaller, structurally intermediary cluster comprising *standards*, *Industry 4.0/5.0*, *adaptation models*, and *security*. It sits between the fabrication core and the supply chain cluster, though it lacks strong ties to either.

**Cluster 4 — Power Electronics and Efficiency (Blue):** The most isolated cluster, containing *industrial ecology*, *inverters*, *efficiency*, and *grid control*. Its peripheral position and sparse edges suggest that power electronics and energy efficiency remain decoupled from mainstream semiconductor sustainability literature.

The network reveals two primary structural holes: the fabrication-to-supply-chain bridge is weak, isolating AI-driven optimization from resilience terms, while standards and power electronics remain structurally disconnected. To close these gaps, two integration pathways are proposed. A **grid-to-core pathway** would embed power electronics energy data into virtual metrology loops, grounding CBAM emissions reporting in real-time process metrics. A **standards-through-supply-chain pathway** would route SSbD and emissions requirements upstream through supply chain management into fabrication governance, embedding compliance as a design constraint rather than a downstream auditing burden.

## V. DISCUSSION

Bibliometric evidence suggests current fragmented approaches can be integrated. To transition toward a coherent framework, our three research questions are revisited as components of a unified System of Systems (SoS) challenge.

### *A. Research Questions Revisited*

*RQ1 —Integration and Generalization*: Semiconductor analytics and metrology systems can integrate heterogeneous data by embedding lifecycle governance into federated data spaces, ensuring cross-vendor and cross-node interoperability with ESG frameworks [43], [44], [45]. However, scaling these architectures without compromising traceability remains a significant gap [46], [47].

*RQ2 —Adaptation Under Constraints*: Platforms maintain reliability despite scarce or imbalanced data via fine-tuning, virtual metrology, and distributed pipelines [8], [48], [49], provided a systems-thinking approach is applied to balance risk across constrained environments [43], [44].

*RQ3 —Efficiency and Compliance*: Treating optimization as an SoS challenge—combining model distillation with hybrid governance [50], [49] — aligns throughput gains with AI safety mandates and Scope 3/CBAM compliance [45].

Consequently, semiconductor analytics must adopt an SoS paradigm to prevent suboptimal subsystem outcomes. Achieving compliance requires architectures that balance local operational trade-offs with global lifecycle orchestration, positioning SSbD as a sector-wide systems challenge rather than a collection of discrete technical fixes [44], [51]. The prominent role of EU institutions — IMEC, Infineon, ASML, and Siemens — signals both the capacity and regulatory motivation to drive the standardization of these principles [52], [53]. Safe and sustainable innovation integrates governance and optimization into a unified SoS architecture.

### B. Match and Contribution

Fig. 4 presents a proposed six-layer architecture harmonizing the 2026 European Commission SSbD Stakeholder Agenda [10], [11] with the 2024 IRDS roadmaps. It is intended to stimulate discussion among researchers, standards bodies, and industry practitioners, not as a validated operational model but as a structured conceptual scaffold. Each layer directly addresses the structural holes identified in the bibliometric analysis — the weak fabrication-to-sustainability bridge, the peripheral position of standards and power electronics, and the absence of lifecycle governance connecting process-level data to regulatory accountability.

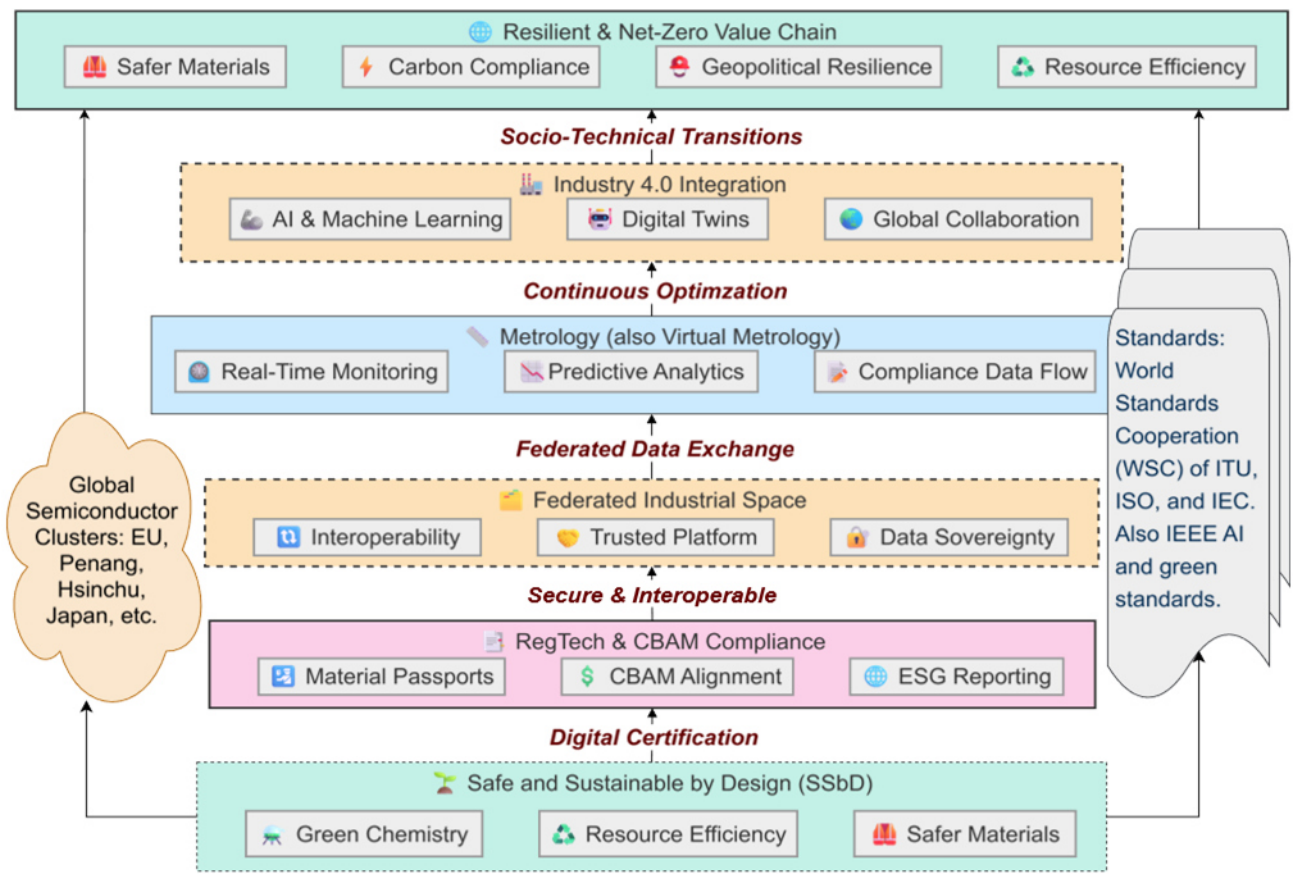

Fig. 4. Layered SSbD architecture for the Semiconductor Industry— a first-iteration of the SSbD Stakeholder Agenda and IRDS roadmap integration.

***SSbD Substitution Layer (Base)***: Grounded in the IRDS ESHS-ESSF roadmap [5], this layer addresses hazardous chemicals (PFCs, PFAS, solvents) [54], [55] and scarce materials (rare earths, gallium, cobalt) by developing safer alternatives [56] and deploying advanced abatement to minimize residual emissions [57]. This directly closes the gap between safety research and downstream compliance.

***RegTech and CBAM Compliance Layer***: This layer operationalizes digital material passports certifying provenance, toxicity, and carbon footprint, aligning CBAM with the IRDS Outside System Connectivity (OSC) roadmap [58] for factory integration and cross-organizational federated trust. Embedding compliance into passports transforms regulatory obligations into interoperable, provenance-aware data flows. Crucially, as generative AI and automated text-screening increasingly govern these pipelines, the layer integrates defensive prompt structures to mitigate "compliance gaming"—preventing supply chain vendors from semantically manipulating internal research reports or safety abstracts to artificially inflate sustainability ratings [59].

***Federated Industrial Spaces Layer***: This resolves the supply-chain structural hole by providing data sovereignty, global standard interoperability [60], distributed governance, and federated machine learning. This enables localized nodes to deploy adaptive, multi-criteria evaluation strategies [61] to bridge global supply chain disparities and achieve compliance without compromising data sovereignty across regions.

***Metrology-Based Optimization Layer***: This layer closes the grid-to-core gap by integrating smart manufacturing [62], virtual metrology [6] and predictive control into closed-loop feedback [63], reducing energy consumption [64], waste and cycle time. Inline sensors and predictive models feed digital twins, material passports, and CBAM reporting systems in the layers above, providing a metrology-grounded single source of truth for ESG sustainability reporting [60].

***Industry 4.0 Layer***: This layer deploys AI to simulate sustainability trade-offs before implementation [65], with Digital Twins extending Virtual Metrology by serving as virtual replicas of semiconductor processes that balance yield against energy, water, and chemical footprints prior to fabrication, thereby linking its outputs directly as compliance data streams. They should ideally enable regulators to coordinate global collaboration through industry consortia.

***Socio-Technical Transitions Layer (Capstone)***: This layer constitutes a self-correcting ecosystem mechanism, —a component which is missing in from the existing literature— where safer materials are validated, compliance is automated, data sovereignty is preserved, and supply chain resilience is achieved through multipolar diversification.

A system walkthrough can be instanced by introducing grid-to-core electronic devices to optimize smart fab energy efficiency. Once a chip is certified via bottom-layer SSbD protocols, its digital provenance is propagated through the RegTech, data exchange, and metrology layers. Ultimately, digital twin simulations validate the cascading, multi-scale impacts of chip adoption before physical scaling.

Together, this layered architecture integrates RegTech, Industry 4.0, and SSbD, bridging the structural holes between fabrication intelligence, sustainability governance, and global standards compliance. By fostering circular material flows, cross-sectoral interoperability, and molecule-to-market traceability, this framework effectively closes the gap between foundational safety research and downstream regulatory accountability, building a safer, net-zero “chemical-to-chip” value chain.

## VI. Conclusion

This scoping review, while limited to indexed bibliometric sources and awaiting empirical validation, charts research trajectories essential for semiconductor excellence. While current streams remain fragmented, the transition toward *Intelligent, Sustainable, and Scalable manufacturing* offers a unified path, showing the need for interoperability, provenance-aware data fabrics, and RegTech integration. No evidence supports a full decoupling of the two major world powers in research; however, domain fragmentation produces suboptimal outcomes, reinforcing the need for systems thinking, across metrology, AI, ESG compliance, and federated governance.

The proposed SSbD architecture (Fig. 4), the main contribution though still an early-stage synthesis, serves as a starting point for community co-development. Its boundaries, sequencing, and governance interfaces are hypotheses to be refined through stakeholder workshops, industrial pilots, and iterative alignment with evolving EC SSbD and IEEE IRDS guidance. By positioning metrology and analytics not as discrete technical fixes but as system-level enablers, the architecture advances intelligent manufacturing that integrates real-time optimization with lifecycle accountability.

Historically, the semiconductor sector has thrived on standardized knowledge transfer — exemplified by "Copy Exactly!" and, the TSMC-led foundry model and compliance with SEMI standards— transforming into a global laboratory where codified knowledge is rapidly shared. This tradition must now extend to sustainability governance through pioneering initiatives. Aligning with macro IRDS roadmaps, emerging initiatives (e.g., IMEC.netzero) quantify process-level energy and environmental footprints; while ESMC's Dresden fab — a joint venture of TSMC, NXP, Infineon, and Bosch — embeds EU sovereignty and sustainability mandates into advanced-node production. Infineon's Smart Power Fab 2026 integrates circularity as a core design principle, driving both scalable and sustainable manufacturing. EU institutions and industry can jointly drive *Scalable* and *Sustainable Manufacturing*, while expanding IRDS/ISRDS Workshop participation to strengthen the global net-zero value chain.

Beyond efficiency gains, resilient and circular ecosystems require data-driven resource efficiency, water stewardship and supply chain resilience. Strengthening partnerships positions regulatory compliance as a catalyst for innovation. Future research must validate SSbD pathways through industrial pilots, exploring how federated data spaces operationalize system-level resilience and how process-level metrology substantiates CBAM and ESG reporting obligations, aligning high-precision manufacturing with the imperatives of the net-zero transition and global sustainability.